\documentclass{emulateapj}
\usepackage{graphicx}

\singlespace

\shortauthors{Kogut et al.}
\shorttitle{ARCADE Instrument}
\slugcomment{Submitted to The Astrophysical Journal}

\begin{document}

\title{An Instrument to Measure the Temperature 
of the Cosmic Microwave Background Radiation
at Centimeter Wavelengths}

\author{A. Kogut\altaffilmark{1},
D. J. Fixsen\altaffilmark{2},
S. Levin\altaffilmark{3},
M. Limon\altaffilmark{2},
P. M. Lubin\altaffilmark{4},
P. Mirel\altaffilmark{2},
M. Seiffert\altaffilmark{3},
and
E. Wollack\altaffilmark{1}}

\altaffiltext{1}{Code 685, Goddard Space Flight Center, Greenbelt, MD 20771}
\altaffiltext{2}{SSAI, Goddard Space Flight Center, Greenbelt, MD 20771}
\altaffiltext{3}{Jet Propulsion Laboratory, California Institute of Technology,
		4800 Oak Grove Drive, Pasadena, CA 91109}
\altaffiltext{4}{Dept. of Physics, University of California, Santa Barbara, CA}

\email{Alan.J.Kogut@nasa.gov}

\begin{abstract}
The Absolute Radiometer for Cosmology, Astrophysics, and Diffuse Emission
(ARCADE)
is a balloon-borne instrument
to measure the temperature of the cosmic microwave background
at centimeter wavelengths.
ARCADE uses narrow-band cryogenic radiometers
to compare the sky to an external full-aperture calibrator.
To minimize potential sources of systematic error,
ARCADE uses a novel open-aperture design
which maintains the antennas and calibrator
at temperatures near 3 K
at the mouth of an open bucket Dewar,
without windows or other warm objects
between the antennas and the sky.
We discuss the design and performance of the ARCADE instrument
from its 2001 and 2003 flights.
\end{abstract}



\section{Introduction}
The cosmic microwave background is a thermal relic from a hot, dense phase
in the early universe.
Deviations from a perfect blackbody spectrum
carry information
on the energetics of the early universe.
Measurements across the peak of the spectrum
limit deviations from a blackbody
to less than 50 parts per million
\citep{fixsen/etal:1996,gush/etal:1990}.
Direct observational limits at longer wavelengths, though, 
are weak: 
distortions as large as 5\% could exist 
at wavelengths of several centimeters or longer
without violating existing observations.

Plausible physical processes
can generate observable distortions
at long wavelengths
without violating limits established at shorter wavelengths.
The decay of massive particles
produced near the Big Bang
imparts a chemical potential to the CMB,
creating a deficit of photons at long wavelengths
\citep{sunyaev/zeldovich:1970,
silk/stebbins:1983,
burigana/etal:1995,
mcdonald/etal:2001,
hansen/haiman:2004}.
Reionization of the universe
by the first collapsed structures
distorts the spectrum
through thermal bremsstrahlung by the ionized gas,
characterized by a quadratic rise
in temperature at long wavelengths
\citep{bartlett/stebbins:1991}.
Reionization is expected to produce a cosmological free-free background
with amplitude of a few mK at frequency 3 GHz
\citep{haiman/loeb:1997,
oh:1999}.
Such a signal is well below current observational limits,
which only constrain free-free distortions to
$\Delta T < 19 $ mK at 3 GHz
\citep{bersanelli/etal:1994}.

Detecting the signal from reionization
requires accuracy of order 1 mK
at frequencies below 30 GHz.
Coherent receivers 
using High Electron Mobility Transistor (HEMT) amplifiers
can easily achieve this sensitivity,
leaving systematic error as the limiting uncertainty
in previous CMB measurements below 30 GHz
(see \cite{kogut:1992} for an experimental review).
These fall into 3 categories.
Below 2 GHz
the dominant uncertainty is synchrotron emission within the Galaxy.
Ground-based measurements at higher frequencies
are limited by atmospheric emission,
while balloon-borne experiments
have been limited by emission from warm parts of the instrument.
The Absolute Radiometer for Cosmology, Astrophysics, and Diffuse Emission
(ARCADE)
is a fully cryogenic, balloon-borne instrument
designed to avoid these sources of systematic error
and provide new limits on deviations
from a blackbody spectrum at centimeter wavelengths
comparable to the limits established at millimeter wavelengths.

ARCADE represents a long-term effort to
characterize the CMB spectrum
at cm wavelengths
in order to constrain the thermal history of the early universe.
An engineering flight designed to test the cold open optics
launched from Ft Sumner, NM on November 2 2001 UT.
A second flight, optimized for CMB observations,
launched from Palestine, TX on June 15 2003 UT.
Scientific analysis of the 2003 flight is presented by
\citet{fixsen/etal:2004}.
We describe the design of the ARCADE instrument
and discuss its performance 
during the 2001 and 2003 flights.

\begin{figure}
\plotone{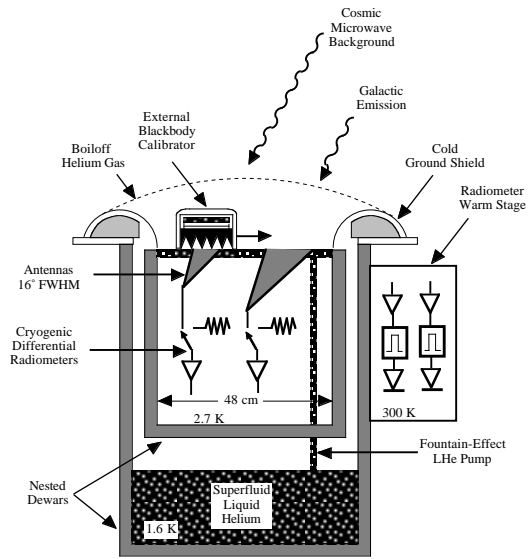}
\caption{ARCADE instrument schematic.
Cryogenic radiometers compare the sky to 
an external blackbody calibrator.
The antennas and external calibrator
are maintained near 2.7 K
at the mouth of an open bucket Dewar;
there are no windows or other warm objects
between the antenna and the sky.
\label{arcade_schematic} }
\end{figure}

\section{Instrument Design}
Figure \ref{arcade_schematic} shows an overview
of the ARCADE instrument design.
It consists of a set of narrow-band cryogenic radiometers
observing at frequencies 10 and 30 GHz
from a balloon payload at 35 km altitude.
Each radiometer measures the difference in power
between a beam-defining corrugated horn antenna
(16\arcdeg ~full width at half maximum)
and a temperature-controlled internal reference load.
An independently controlled external blackbody calibrator
(emissivity $\epsilon > 0.9997$)
covers each antenna aperture in turn,
so that each antenna alternately views
the sky or a known blackbody.
The calibrator, antennas, and radiometers
are maintained near 2.7 K
at the mouth of an open bucket Dewar
with no windows or other warm object between the antennas and the sky.
Fountain-effect pumps lift superfluid liquid helium
to reservoirs at the antenna aperture plane
and inside the external calibrator
to maintain cryogenic temperatures.
Boiloff helium gas from the main Dewar tank
vents through the aperture plane
to provide additional cooling
and prevent the condensation of atmospheric nitrogen
on the optics.

\begin{figure}[b]
\plotone{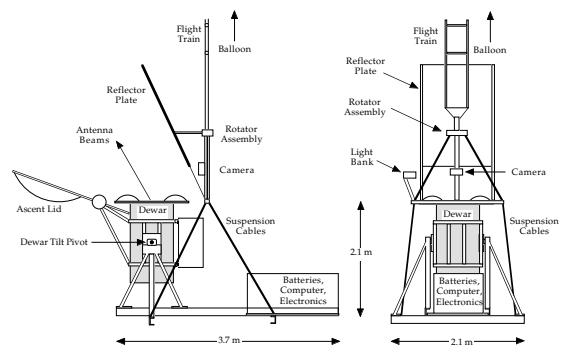}
\caption{ARCADE gondola in 2003 flight configuration.
A deployable lid protects the cold optics during launch and ascent.
The antennas view the sky 30\arcdeg ~from the zenith.
The reflective shield screening the flight train and parachute
from view of the antennas
is the largest single source of systematic uncertainty
and is measured in flight by tipping the dewar.
\label{agondola_schematic} }
\end{figure}

Figure \ref{agondola_schematic} shows the entire payload.
The Dewar is mounted on a gondola
suspended 64~m below the balloon.
Batteries and flight electronics 
mounted on a movable pallet at the other end of the gondola from the Dewar
serve as counterweights,
allowing the suspension to be located 
well away from the antenna apertures.
A deployable lid covers the optics during launch and ascent
and can be commanded open or closed at float.
The antennas view the sky 30\arcdeg ~from the zenith
to avoid direct view of the balloon or flight train.
A metallized foam reflector plate
attached to the suspension
shields the flight train, parachute, 
termination package,
and part of the balloon
from direct view of the antennas.
The dewar can be commanded to tip at angles up to 25\arcdeg
~from vertical,
changing the orientation of the antennas
with respect to the balloon and reflector.
The electronics pallet can also be moved in flight
to trim the gondola pitch.
During nominal operation
the dewar is vertical
while the gondola rotates at approximately 0.5 RPM
to scan the antennas around a ring 30\arcdeg ~from the zenith.
No pointing control is required;
magnetometers and inclinometers
allow pointing reconstruction within 3\arcdeg.
A Global Positioning System receiver
collects position and altitude information.
Table \ref{gondola_summary_table} summarizes the payload.

\subsection{Radiometers}
\begin{figure}[t]
\plotone{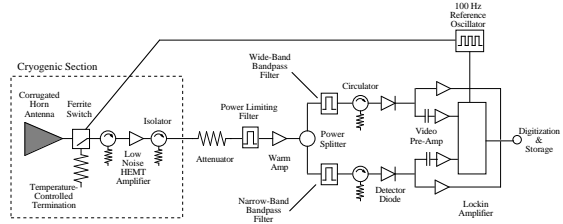}
\caption{Block diagram of the ARCADE radiometers.
\label{block_diagram} }
\end{figure}

Figure \ref{block_diagram} shows a block diagram of the radiometers.
Each radiometer uses a cryogenic 
HEMT-based front-end amplifier.
They are
switched at 100 Hz between a corrugated conical horn antenna
and a temperature-controlled waveguide load
using latching ferrite waveguide switches
with measured insertion loss below 0.4 dB.
The low insertion loss greatly reduces requirements for
thermal monitoring and control of the instrument front end.
All of the radiometer back ends are housed 
in a temperature-controlled module mounted to the outside of the dewar,
with stainless steel waveguide (30 GHz) or coax (10 GHz)
providing the RF link between the cryogenic and room-temperature components.
The back end of each radiometer is split into two frequency sub-channels:
a wide-band channel for maximum sensitivity,
and a narrow-band channel restricted to protected RF bands.
Video preamplifiers following each detector diode
separately amplify the dc 
and ac portions of the signal,
proportional to the
total power on the diode
and the antenna-load temperature difference, respectively.
A lockin amplifier demodulates the switched (ac) signal
and integrates for one second
to produce an output proportional to the difference in power
between the antenna and the internal load.
The lockin output, total power,
and assorted housekeeping voltages
are digitized at 1 Hz
and stored in an on-board computer.
Table \ref{rad_table} summarizes the radiometer properties.

The internal reference load has a dual purpose:
it provides a stable cold reference for the fast gain chop,
and can be adjusted in temperature during flight
to eliminate the radiometric offset 
resulting from imbalance 
in the switch or the two arms of the radiometer.
The reference load
consists of a waveguide termination
mounted on a temperature-controlled plate.
Temperature control and readout consist of 
individually calibrated ruthenium oxide thermistors
in a digitally controlled proportional-integral-differential loop.

\subsection{Antennas And Beam Pattern}
Each receiver is fed by a corrugated conical horn antenna
scaled in wavelength to produce identical beam shapes
at 10 and 30 GHz.
To avoid convective instabilities in the helium vapor barrier,
the instrument should remain nearly vertical during observations.
We reconcile this requirement 
with the need for sky coverage (and avoiding direct view of the balloon)
by mounting the antennas at a 30\arcdeg ~angle from the zenith,
slicing each antenna at the aperture plane.
Quarter-wave chokes surround the resulting elliptical aperture
and provide further suppression of side lobes.
Stainless steel flares at the dewar rim 
function as a cold (20 K) ground shield
to block emission from the Earth.  

\begin{figure}
\includegraphics[angle=90,width=3.25in]{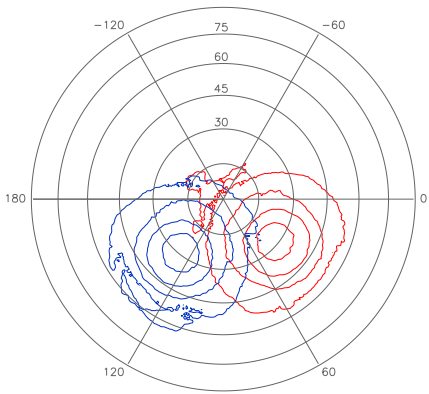}
\caption{Beam patterns measured in flight configuration
at 10.1 GHz (red)
and 30.3 GHz (blue).
The upper $2\pi$ sr visible above the instrument aperture plane
is shown in a stereographic projection,
with the zenith at the center and the horizon at the edge.
Azimuth 90\arcdeg ~points along the gondola long axis
away from the gondola suspension and flight train.
Contours are -3, -10, -20, and -30 dB.
Despite the sliced antennas,
the beam patterns remain circular past -10 dB.
Some steepening is visible toward the cold flares at low elevation.
\label{beam_pattern} }
\end{figure}

We measured the co-polar and cross-polar beam patterns
of each radiometer
over $4\pi$ sr
at the Goddard Electro-Magnetic Anechoic Chamber.
Beam mapping used the entire dewar in flight configuration,
including the flares and external calibrator.
Figure \ref{beam_pattern}
shows the co-polar beam patterns at 10.1 and 30.3 GHz
in a stereographic projection centered on the dewar aperture plane.
The beam centroids are located 30\arcdeg ~from the zenith
with azimuthal separation 90\arcdeg.
Despite the sliced geometry,
the beams retain circular symmetry past -10 dB,
with full width at half maximum 16\arcdeg ~at both 10 and 30 GHz.
The deformation evident at -30 dB towards low elevation angles
shows the effect of the nearby cold flares.

\begin{table}[t]
\begin{center}
\caption{\label{gondola_summary_table}
ARCADE Payload Summary}
\begin{tabular}{l c c }
\tableline
Property & 2001 & 2003\\
\tableline
Mass\tablenotemark{a} (kg) & 	1148	&	1183 	\\	
Mean Power (W)		&	326	&	325	\\
Peak Power (W)  	&	470	&	925	\\
Float Altitude (km)	&	29.5	&	34.7	\\
Float Pressure (Torr)	&	10	&	4.5 	\\
\tableline
\end{tabular}
\tablenotetext{a}{Mass at launch, excluding ballast}
\end{center}
\end{table}

\begin{figure}[b]
\includegraphics[angle=90,width=3.25in]{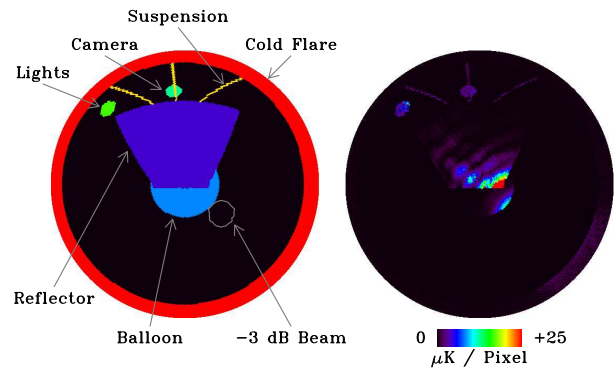}
\caption{Schematic representation of objects included
in model of 10 GHz flight train emission.
(left) Stereographic projection of upper 2$\pi$ sr
viewed from the 10 GHz antenna,
using the same coordinate convention as Figure \ref{beam_pattern}.
The zenith is at the center
while the horizon is at the edge.
Objects are color-coded for visual clarity;
the -3 dB contour for the 10 GHz beam is also shown.
The reflector hides much of the flight train
but only partially screens the balloon.
(right) Antenna temperature contribution from each element,
estimated by convolving the position, emissivity, and temperature
of objects in the left panel
with the measured 10 GHz beam pattern.
Units are flux density in $\mu$K per $1\arcdeg ~\times ~1\arcdeg$ pixel,
so that the total antenna temperature is given by
the sum over all pixels in the figure.
Emission from the reflector dominates the total
but is simple to model.}
\label{flight_train_fig}
\end{figure}

Warm objects in the beam
contribute to the measured sky signal.
We reduce the contribution from the flight train
by mounting a reflective plate to the gondola suspension
and extending over the Dewar aperture.
The reflector plate consists of metallized foam
and blocks direct view of the
rotator assembly,
ladder,
parachute,
and termination package
while partially obscuring the balloon.
We model emission from the suspension, reflector, and balloon
by convolving the temperature, emissivity, and position
of these objects
with the measured beam patterns.
Figure \ref{flight_train_fig} shows the relevant geometry.
The reflector re-directs the part of the beams
that would otherwise view the flight train
to blank regions of the sky.
Emission from the plate is comparable 
to emission from the flight train hidden behind it,
but is much simpler to model.
Table \ref{flight_train_table} summarizes the contribution of each element
as a function of Dewar tip angle.
In normal operation with the Dewar vertical,
warm parts of the instrument contribute a predicted
10.8 mK to the sky temperature.
Tipping the Dewar back toward the electronics pallet
moves the beams closer to the balloon and reflector,
while tipping forward reduces the signal.
The predicted signal modulation
can be observed in flight
\citep{fixsen/etal:2004}.

\subsection{External Calibrator}
The determination of the sky temperature
rests on the comparison of the sky
to an external blackbody calibrator,
described in \cite{kogut/etal:2004}.
The external calibrator consists of
a microwave absorber
(Eccosorb CR-112, an iron-loaded epoxy)
cast with grooves in the front surface to reduce reflections.
The Eccosorb is mounted on a thermal buffer
consisting of
an alternating series of thin copper and fiberglass plates.
Thermal control is achieved by heating the outermost copper plate,
which is in weak thermal contact
with a superfluid helium reservoir inside the target housing.
Fountain-effect pumps lift liquid helium
from the main Dewar to the calibrator.
Reflective multilayer insulation surrounds the calibrator housing.
The calibrator is mounted on a rotating mechanism
to allow it to completely cover
either antenna aperture
while the other antenna views the sky.
The antennas view the calibrator 30\arcdeg ~from normal incidence.
Measurements with the calibrator over each antenna
limit the in-band power reflection coefficient to below -35 dB.
We monitor the calibrator temperature 
at 7 locations
within the Eccosorb and on the copper thermal control plate.
Data taken {\it it situ}
with the target immersed in liquid helium
during 8 independent calibrations spaced 4 years apart
agree within 2 mK,
providing an independent verification of the thermometer calibration
at temperatures near that of the cosmic microwave background.

\begin{table}[b]
\caption{\label{rad_table}
ARCADE Receiver Summary}
\begin{tabular}{l c c c c c c}
\tableline
 & 10 Wide & 10 Narrow & 30 Wide & 30 Narrow \\
\tableline
Band Pass (GHz)			& 9.5 - 10.6 & 10.6 - 10.7 & 28.4 - 31.3 & 31.3 - 32.3 \\
Switch Isolation (dB)		& $>$25      & $>$25       & $>$25       & $>$25 \\
Switch Insertion Loss (dB) 	& 0.4 & 0.3 & 0.3 & 0.3 \\
Cold Stage Gain (dB)		& 36  & 35  & 28  & 26 \\
Total RF Gain (dB)		& 63  & 74  & 64  & 74 \\
System Temperature (K)		& 12  & 12  & 77  & 77 \\
Sensitivity (mK Hz$^{-1/2}$) 	& 0.7 & 2.3 & 2.8 & 4.8 \\
Beam FWHM (degrees)		& 16  & 16  & 16  & 16 \\
\tableline
\end{tabular}
\end{table}

\subsection{Gondola Electronics}
The gondola electronics consist of a flight computer, telemetry 
system, magnetometer-inclinometer box, drive electronics for the 
various motors, analog interface unit, and central electronics box.
The central electronics box houses the ambient temperature and 
pressure readout electronics, voltage and current measuring modules, 
relay boards and the power distribution system, and various other 
electronics. The analog interface unit contains 48 16-bit A/D 
channels and 12 12-bit D/A channels. This unit communicates with the 
flight computer which consists of a i486-based embedded computer 
running a multi-tasking, real-time operating system. 
The computer assembly contains a hard disk 
capable of storing an entire flight's worth of data. 
The flight computer 
communicates with the cryogenic temperature sensor readout electronics 
\citep{fixsen/etal:2002}
via an asynchronous serial (RS-232) interface. 
Flight data are stored on-board and relayed to the ground using a
discriminator channel on the
Consolidated Instrument Package (CIP) 
provided by the National Scientific Balloon Facility (NSBF).
Commands are sent to the flight computer via our 408 MHz 
telemetry uplink that allows 4800 baud ASCII data communication.

A video camera mounted on the reflector above the Dewar aperture
allows direct imaging of the cold optics in flight.
Two banks of light-emitting diodes provide the necessary illumination.
Ground tests revealed interference from the video transmitter
in the 10 GHz wide channel.
We do not use data from any channel for science analysis
during times when the camera and video transmitter are turned on.

\section{Cryogenic Design And Performance}
ARCADE uses cryogenic optics and calibrator
on a 48-cm diameter clear aperture
operating without windows
at temperatures near 2.7 K.
To eliminate any spillover of the antenna beams
onto warm portions of the Dewar walls,
the antenna apertures and external calibrator
are located at the mouth of an open bucket Dewar.
Two fountain-effect superfluid pumps
move helium from the main Dewar into a reservoir
located inside the calibrator;
a third pump lifts liquid helium to cool the aperture plane.
Boiloff gas from the 220 liter main tank
vents through pinholes 
in the metal mounting plate holding the antenna apertures,
providing additional cooling through the enthalpy of the helium gas
while providing a buffer between the optics
and the atmosphere.

\subsection{Thermometry}
We monitor cryogenic temperatures
using 4-wire ac resistance measurements
of 27 ruthenium oxide thermometers
read out every 1.067 seconds
\citep{fixsen/etal:2002}. 
Seven of the thermometers are located in the external calibrator.
Eight additional thermometers
monitor critical temperatures 
(antenna throat,
internal load,
Dicke switch,
and cryogenic HEMT amplifier)
on each of the 2 radiometers.
The remaining thermometers
monitor the temperature
of the aperture plane,
cold flares,
superfluid pumps,
liquid helium reservoirs
(main tank, aperture plane, and external calibrator)
as well as the Dewar walls
and the instrument support structure inside the Dewar.

The in-flight thermometry 
compares the resistance of each thermometer
to a set of 4 calibration resistors
spanning the dynamic range of the thermometer resistances.
The calibration resistors are part of the readout electronics board,
located in a temperature-controlled enclosure.
One of the calibration resistors 
includes the electrical harness into the Dewar
to monitor possible effects from electrical pickup
or stray capacitance.
The flight computer uses a lookup table
to infer a temperature for each thermometer
using temperature-resistance curves
derived from ground tests.
The instrument is mainly a transfer standard
to compare the sky
to the external calibrator target;
precise knowledge of the absolute temperature
is required only for the external calibrator.
The resistance to temperature calibration for
the thermometers in the external calibrator
was obtained multiple times over several years
by measuring the resistance of the embedded thermometers
with the calibrator submerged in liquid helium,
as the pressure over the bath was slowly lowered.
Observations of the superfluid helium transition at 2.1768 K
using the flight readout electronics
agree within 0.3 mK over multiple calibrations spaced 4 years apart,
providing a cross-check on the absolute thermometry.
Thermal gradients during the calibration process 
degrade the absolute accuracy to 2 mK at temperatures near 2.7 K.

\subsection{Heat Flow}
The ARCADE instrument is mounted
at the mouth of an open fiberglass/aluminum composite
bucket Dewar
manufactured by Precision Cryogenics.
The inner diameter is 61 cm
with a fiberglass top section extending 38 cm
to reduce heat flow from the aluminum body.
The aluminum bottom section holds 220 liters of liquid helium.
The heat leak to the cryogen
is dominated by thermal conduction down the walls.
Typical boiloff rates
are 12 liters hr$^{-1}$ on the ground with the ascent lid closed,
32 liters hr$^{-1}$ during ascent,
and 15 liters hr$^{-1}$ at float
when we activate a boiloff heater in the main reservoir
to provide additional outflow of cold gas.
The Dewar is unpressurized
and vents through the aperture plane
so the liquid remains near ambient pressure.
The 2003 flight
launched with 140 liters of liquid helium
in the main reservoir,
falling to 53 liters at float
to provide 3.5 hours of cryogenic observations.

Fountain-effect pumps move superfluid liquid helium
from the main reservoir to separate reservoirs
below the aperture plane
and in the external calibrator housing.
Each pump consists of a 1.3 cm diameter porous plug
(CoorsTek P-1/2-BC ceramic disk
with pore diameter less than 0.5 $\mu$m)
mounted in a leak-tight housing.
A resistive heater
warms the liquid inside the housing 80 mK
above the 1.56 K bath temperature,
creating enough pressure (35 Pa)
to lift the liquid helium
130 cm to the reservoirs at the top of the Dewar.
Two pumps (a primary and a backup) service the target
with a third pump allocated to the aperture plane.
Each pump can lift approximately 2.8 liters hr$^{-1}$ liquid helium,
providing 2.6 W cooling each through the latent heat of the liquid.
Boiloff gas from the main reservoir
provides additional cooling to the aperture plane
through the enthalpy of the cold gas.

\subsection {Atmospheric Condensation}
The cryogenic performance of the ARCADE 
depends on interactions between ambient nitrogen
and the cold helium efflux
at pressures below 5 Torr.
Gaseous helium is denser than the ambient nitrogen
at float altitude
provided the helium remains colder than 20 K.
The mixing as the helium warms and the nitrogen cools
is complicated and difficult to simulate
in pressure chambers on the ground,
which are not designed to maintain
a dilute nitrogen atmosphere
in the presence of a large (12 m$^3$ hr$^{-1}$)
helium gas source.

\begin{table}[t]
\caption{\label{flight_train_table}
Model Flight Train Emission at 10 GHz (mK)}
\begin{center}
\begin{tabular}{l c c c c c}
\tableline
Component & \multicolumn{5}{c}{Dewar Tip Angle\tablenotemark{a}} \\
          &     -10     &  -5    &    0    &    5    &   10   \\
\tableline
    Balloon  &	 2.1	&  1.4   &   0.7   &   0.2   &	 0.1  \\
  Reflector  &	25.6	& 10.6   &   4.8   &   2.4   &   1.3  \\
     Camera  &	 0.4	&  0.2   &   0.2   &   0.1   &	 0.1  \\
     Lights  &	 0.2	&  0.2   &   0.3   &   0.3   &	 0.2  \\
 Suspension  &	 4.7	&  3.9   &   3.2   &   2.5   &	 1.8  \\
Cold Flares  &	 1.4	&  1.4   &   1.4   &   1.4   &	 1.4  \\
     Ground  &	 0.3	&  0.3   &   0.4   &   0.4   &	 0.4  \\
\tableline
Total        &  34.7    & 18.1   &  10.8   &   7.4   &	 5.3  \\
\tableline
\end{tabular}
\tablenotetext{a}{Positive angles tip the Dewar forward
and the antenna beams away from the zenith.}
\end{center}
\end{table}

The efflux of helium gas above the aperture
does not eliminate condensation of atmospheric nitrogen
onto the cryogenic optics,
but does reduce it to levels compatible with 
the desired CMB observations.
Flight data show excess heat dissipation on the aperture plane,
consistent with an accumulation rate of approximately 
200 g hr$^{-1}$ of nitrogen onto the aperture,
a heat source of order 6 W.
Visual examination using the video camera
confirms this slow accumulation of nitrogen ice,
with several mm of ``frost'' visible on the optics
an hour after opening the ascent lid.
Nitrogen ice is nearly transparent at cm wavelengths;
since the ice cools to the same temperature as the rest of the optics,
modest accumulations primarily affect only mechanical operations.
The 2003 flight included a large (kW) heater on the cold optics,
including the antenna aperture and throat sections.
With the superfluid pumps off,
the aperture plane and calibrator have only weak thermal coupling to the Dewar.
Raising the aperture plane and calibrator above 80 K for 15 minutes
suffices to remove accumulated nitrogen ice
without only a modest impact on the main liquid helium reservoir.

\section{Discussion}
Both the 2001 and 2003 flights 
demonstrate the viability of 
large open-aperture cryogenic optics
for CMB measurements.
The ARCADE cryogenic design
maintains the external calibrator,
antennas,
and cryogenic radiometers
at temperatures near 2.7 K
for many hours at 35 km altitude.
Cold boiloff gas vented through the aperture plane
reduces condensation of atmospheric nitrogen
to levels consistent with the desired CMB observations;
the primary impact is the cooling required to handle
the additional heat load on the aperture.
The current instrument is relatively small
(two frequency channels at 10 and 30 GHz)
and is intended primarily as a pathfinder
to verify the cryogenic design
under flight conditions.
A larger second-generation instrument
with 6 channels exending down to 3 GHz
is currently under construction
and is scheduled to launch in 2005.

\acknowledgements
We thank the NSBF staff in Palestine and Ft Sumner
for their capable support throughout integration, launch, flight, and recovery.
We thank M. DiPirro and D. McHugh of the Cryogenic Fluids Branch
at GSFC for supporting the ARCADE design and thermometer calibration.
Interns Gary Palmer II and Anatoly Brekhman
made many of the thermometers used in the 2003 flight.
Rachel Maire and Alexander Rischard assisted payload integration.
Leah Johnson provided
valuable assistance for the flight electronics.
Elizabeth Cantando aided cryogenic integration
and provided flight support.
Doron Halevi and Peter Meinhold integrated RF communications.
ARCADE is supported by the National Aeronautics and Space Administration
under the Space Astrophysics and Research Analysis program
of the Office of Space Science.
The research described in this paper was carried out in part at 
the Jet Propulsion Laboratory, California Institute of Technology,
under contract with the National Aeronautics and Space Administration.

\bibliographystyle{apj}
\bibliography{arcade_instrument}

\begin{thebibliography}{15}
\expandafter\ifx\csname natexlab\endcsname\relax\def\natexlab#1{#1}\fi

\bibitem[{{Bartlett} \& {Stebbins}(1991)}]{bartlett/stebbins:1991}
{Bartlett}, J.~G. \& {Stebbins}, A. 1991, \apj, 371, 8

\bibitem[{{Bersanelli} {et~al.}(1994){Bersanelli}, {Bensadoun}, {de Amici},
  {Levin}, {Limon}, {Smoot}, \& {Vinje}}]{bersanelli/etal:1994}
{Bersanelli}, M., {Bensadoun}, M., {de Amici}, G., {Levin}, S., {Limon}, M.,
  {Smoot}, G.~F., \& {Vinje}, W. 1994, \apj, 424, 517

\bibitem[{{Burigana} {et~al.}(1995){Burigana}, {De Zotti}, \&
  {Danese}}]{burigana/etal:1995}
{Burigana}, C., {De Zotti}, G., \& {Danese}, L. 1995, \aa, 303, 323

\bibitem[{{Fixsen} {et~al.}(1996){Fixsen}, {Cheng}, {Gales}, {Mather},
  {Shafer}, \& {Wright}}]{fixsen/etal:1996}
{Fixsen}, D.~J., {Cheng}, E.~S., {Gales}, J.~M., {Mather}, J.~C., {Shafer},
  R.~A., \& {Wright}, E.~L. 1996, \apj, 473, 576

\bibitem[{{Fixsen} {et~al.}(2004){Fixsen}, {Kogut}, {Levin}, {Limon}, {Lubin},
  {Mirel}, {Seiffert}, \& {Wollack}}]{fixsen/etal:2004}
{Fixsen}, D.~J., {Kogut}, A., {Levin}, S., {Limon}, M., {Lubin}, P.~M.,
  {Mirel}, P., {Seiffert}, M., \& {Wollack}, E. 2004, \apj, submitted

\bibitem[{{Fixsen} {et~al.}(2002){Fixsen}, {Mirel}, {Kogut}, \&
  {Seiffert}}]{fixsen/etal:2002}
{Fixsen}, D.~J., {Mirel}, P.~G.~A., {Kogut}, A., \& {Seiffert}, M. 2002, Rev.
  Sci. Inst., 73, 3659

\bibitem[{{Gush} {et~al.}(1990){Gush}, {Halpern}, \&
  {Wishnow}}]{gush/etal:1990}
{Gush}, H.~P., {Halpern}, M., \& {Wishnow}, E.~H. 1990, \prl, 65, 537

\bibitem[{{Haiman} \& {Loeb}(1997)}]{haiman/loeb:1997}
{Haiman}, Z. \& {Loeb}, A. 1997, \apj, 483, 21

\bibitem[{{Hansen} \& {Haiman}(2004)}]{hansen/haiman:2004}
{Hansen}, S.~H. \& {Haiman}, Z. 2004, \apj, 600, 26

\bibitem[{{Kogut}(1992)}]{kogut:1992}
{Kogut}, A. 1992, in {Current Topics in Astrofundamental Physics}, ed.
  N.~{Sanchez} \& A.~{Zichichi} ({Singapore}: {World Scientific}), 137

\bibitem[{{Kogut} {et~al.}(2004){Kogut}, {Fixsen}, {Levin}, {Limon}, {Lubin},
  {Mirel}, {Seiffert}, \& {Wollack}}]{kogut/etal:2004}
{Kogut}, A., {Fixsen}, D.~J., {Levin}, S., {Limon}, M., {Lubin}, P.~M.,
  {Mirel}, P., {Seiffert}, M., \& {Wollack}, E. 2004, Rev. Sci. Inst.,
  submitted

\bibitem[{{McDonald} {et~al.}(2001){McDonald}, {Scherrer}, \&
  {Walker}}]{mcdonald/etal:2001}
{McDonald}, P., {Scherrer}, R.~J., \& {Walker}, T.~P. 2001, \prd, 63, 023001

\bibitem[{{Oh}(1999)}]{oh:1999}
{Oh}, S.~P. 1999, \apj, 527, 16

\bibitem[{{Silk} \& {Stebbins}(1983)}]{silk/stebbins:1983}
{Silk}, J. \& {Stebbins}, A. 1983, \apj, 269, 1

\bibitem[{{Sunyaev} \& {Zel\'dovich}(1970)}]{sunyaev/zeldovich:1970}
{Sunyaev}, R.~A. \& {Zel\'dovich}, Y.~B. 1970, Ap. Space Sci., 7, 20

\end{thebibliography}

\end{document}